# Effect of eccentric mixing parameters on chaotic characteristics and mixing time for viscous liquid based on sound decibels

Ronfgang Wang[1], Lijun Zhao[1,2], Yunshi Yao[1,2]

1.Key Laboratory of Highway Construction Technology and Equipment of Ministry of Education, School of Construction Machinery, Chang'an University, Xi'an 710064, China

2.Detong Intelligent Technology Co., Ltd., Xuchang 4641000, China

Correspondence:yaoys@chd.edu.cn

**Abstract:** Eccentric mixing is a typical chaotic mixing method, and the study of its mixing characteristics is beneficial to the optimization of the mixing process. In this study, the effects of eccentricity (E/R) and rotational speed (N) on the mixing time are quantified through tracer staining experiments and image grayscale analysis, and the method of calculating the Lyapunov exponent (LLE) based on the sound decibel value time series is proposed. Experiments show that sound decibel value time series can better characterize the chaotic dynamics of the system. Based on the control variable method, the mixing time and LLE all show a nonlinear trend of decreasing and then increasing, or increasing and then decreasing, with the increase of eccentricity and rotational speed. When E/R=0.4 and the rotational speed N=450rpm, the mixing time was shortened by 48% compared with the center mixing, and the degree of chaos reached the peak. The dimensionless chaos indicator model $M = 22200(\frac{E}{R})^{-0.17}(\frac{\rho ND^2}{\mu})^{-3.14}$ is further constructed to reveal the quantitative relationship between the eccentric stirring parameters and the chaos intensity. This method provides a theoretical basis for the real-time monitoring of the chaotic characteristics of complex flow fields and the optimization of industrial mixing equipment.

## 1.Introduction

Mixing technology is a common operation unit to enhance mass transfer and reaction efficiency in multiphase systems, is wide used in many fields, such as petrochemical, biopharmaceutical and environmental treatment. The structural stability, porosity, and fluidity of process products such as chemical synthesis and ceramic or concrete preparation are all affected by mixing homogeneity. In traditional center mixing, the material force and movement form are limited by the symmetric flow field. Problems such as obvious energy attenuation and wide low-efficiency stirring zone are widespread. Material agglomeration is so serious that mixing efficiency is low. Chaotic mixing technology disrupts the symmetry of the flow field through irregular disturbances, based on nonlinear dynamic theory. Chaotic phenomenon inducing the material to undergo the complex dynamics of stretching-folding-reorientation behavior in mixing tank, the mixing interface evolution and deformation acceleration, the mixing results are better [1-4]. Deyin Gu team used the rigid-flexible impeller coupled with the chaotic motor found that the larger the chaotic index of the mixing tank, the more uniformly dispersed the solids in the stirred tank[5].

Eccentric stirring asymmetrizes the flow field structure by changing the spatial position of the stirring shaft [6-9]. Existing studies have shown that eccentric stirring parameters have an important effect on the multiphase mixing performance. DeyinGu [10] found that eccentric stirring combined with a misaligned fractal impeller can disrupt the symmetry of the flow field and improve the axial circulation efficiency and dispersion uniformity of floating particles through CFD numerical simulations. Mengxue Zhang et al. investigated by using the planar laser-induced fluorescence (PLIF) technique and found that eccentric stirring is an effective way to enhance the viscous liquid mixing[11].

At the level of experimental study, the transparent tank visualization technique provides an intuitive observation means for the flow field analysis. Wenjie Li team et al. used fluorescence tracer method to study the enhancement of purification stirring by eccentric impeller stirring in cadmium removal process, verified the simulated flow field mixing results, and explored the fluid's motion law[12]. Zhangjiaqi [13] used the flow field visualization experiment and the computational fluid dynamics simulation to verify that the axial flow is the main reason for the different mixing efficiencies between the same direction and the opposite direction in the planetary motion of the double stirring paddles with the effect of stretching, shearing, and folding of the fluid. Mingyang Fan, Jianxin et al. injected $Na_2S_2O_3$-$5H_2O$ solution into a transparent cylinder to investigate the mixing efficiency of a reciprocating-rotating coupled stirrer, and the mixing time was digitized during the mixing process by using the sensory hash algorithm[1]. Y. Hirata calculated the power requirement for the reciprocating motion of the disk, and the mixing time was determined using the decolorization reaction of iodine with sodium thiosulfate, and found that the chaotic mixing field achieves a faster mixing than the normal mixing[14]. Above all traditional mixing time detection methods (conductivity method and pH tracer method) rely on invasive sensors, which are costly and susceptible to interference by fluid properties. Although the hash algorithm and decolorization reaction can distinguish the liquid staining changes, the judgment process is subjective, the color perception is not sensitive enough, and the quantification accuracy is insufficient.

Chaos indicators are used to quantify the chaotic properties of nonlinear dynamical systems, and common chaos indicators include Lyapunov exponent (LLE), fractal dimension, etc., of which LLE utilizes the particle separation velocity to study system chaos. In the existing mixing process, the calculation of LLE using time series such as power and torque is the mainstream method, but it is prone to the problems of insufficient short-time data collection and insufficient data of kinetic characteristics[15].

To address the above problems, this study utilizes the dye tracer method to visualize the experiments and low-cost image grayscale recognition to ensure mixing time nodes, collects the sound decibel time series during solid-liquid mixing to study the relationship between LLE and eccentricity, rotational speed, and mixing time, and utilizes the mixing system parameter dimensionless chaotic dynamics indexes.

Table1  Explanation of symbol

| Nomenclature | | $\lambda$ | Maximum Lyapunov Index |
|---|---|---|---|
| e | Eccentricity of the mixing shaft(mm) | D | Diameter of mixing paddle(mm) |
| E | Eccentricity of the mixing shaft(mm) | N | Mixing shaft speed(rpm) |
| R | Effective mixing radius of mixing paddle blade(mm) | $\rho$ | liquid density $(kg \cdot m^{-3})$ |
| $\frac{E}{R}$ or $\frac{e}{R}$ | Eccentricity of the mixing shaft in relation to the mixing cylinder | $\mu$ | Fluid Power Viscosity $(Pa \cdot S)$ |
| T | Diameter of mixing tank(mm) | | |

## 2.Theoretical Foundations

2.1 Chaos metric: Lyapunov exponent

The Lyapunov exponent (LLE) measures the separation rate of particles in neighboring orbits in a dynamical system. The mathematical model of Lyapunov exponent is shown in Eq. 1, where $\delta(t)$ is the distance between two neighboring orbits after time t. The LLE is a measure of the separation rate of particles in neighboring orbits in a dynamical system.

$$\lambda = \lim_{t \to \infty} \lim_{\delta(0) \to 0} \frac{1}{t} \ln \frac{|\delta(t)|}{|\delta(0)|} \quad (1)$$

Maximum Lyapunov exponent is an important parameter for evaluating the dynamic properties and initial value sensitivity of chaotic systems. LLE>0 indicates that the adjacent orbits in the phase space are exponentially separated, that is, the initial value sensitivity. The larger the LLE is, the stronger the degree of chaos is. In this work, the virtual environment of matlab was embedded in python, and the framework of Wolf algorithm was constructed to calculate the maximum Lyapunov exponent $\lambda$. The core principle of Wolf algorithm was to estimate the LLE by tracking the dispersion rate of the neighboring orbits in the phase space. According to Takens' embedding theorem, the one-dimensional time series was mapped to the high-dimensional phase space, and the topological structure of the original dynamical system was recovered. The time series x (t) is reconstructed into the phase space modeled as

$$X(t) = [X(t), X(t+\tau), X(t+2\tau), ..., X(t+(m-1)\tau)], m \geq 2d+1$$

where d is the system dimension, m is the embedding dimension that ensures that no orbital overlap occurs in the reconstructed phase space, and $\tau$ is the delay time that balances the independence and correlation between data points. In addition, the number of neighbors K and the number of iterations s are also important physical parameters to ensure the fidelity of the Wolf algorithm.

2.2 Theoretical derivation of the degree of chaos and parameter relationships

Stirring tanks are generally used for solid-liquid, gas-liquid, and multi-liquid mixing, and the continuity, Navier-Stoikes, and particle diffusion equations are integrated to analyze the mixing characteristics.

Continuity equation is

$$\nabla \cdot v = 0$$

Navier-Stoikes equation is

$$\rho(\frac{\partial v}{\partial t} + v \cdot \nabla v) = -\nabla p + \nabla \cdot \tau + f$$

where $\tau$ is the stress tensor; $f$ is the volumetric force (external force acting on a unit volume of the fluid); $\rho$ represents the density of the fluid; and $p$ is the pressure.

Particle diffusion equation is

$$\frac{\partial c}{\partial t} + v \cdot \nabla c = D\nabla^2 c + s$$

where $c$ is the particle concentration, $D$ is the particle diffusion coefficient, and $s$ is the source term.

It can be seen that changes in the velocity field $v$ are related to the fluid forces $f$, properties $\rho$, and diffusion of the moving particles $D$. The velocity field changes help to complicate the diffusion of the particles and the change of the stress tensor at the fluid contact surface, and increase the trajectory traversability of the particles and chaotic properties such as tensile deformation of the fluid interface. Therefore, the dimensionless parameter of the degree of chaos in the mixing system must include the mixing parameter $N$ and the eccentricity $e$, as well as the material properties.

## 3. Experimental

3.1 Laboratory Instruments

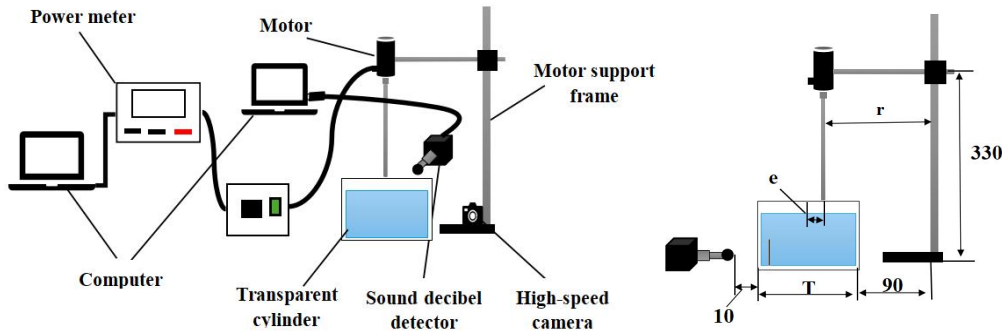

Figure 1. Desktop simple visualization of the experimental equipment diagram and dimensions

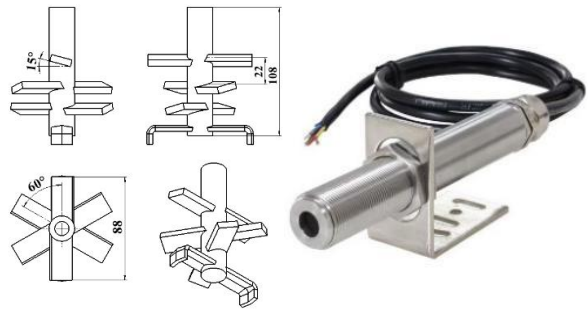

Figure 2. Stirring paddles and sound decibel tester

The total equipment for the experiment is shown in Figure 1, the experiment is carried out in a transparent acrylic cylindrical cylinder with a cylinder diameter of 200mm (T), wall thickness of 2mm, and a height of 90mm, the structure and dimensions of the stirring paddles are shown in Figure 2, the stirring paddles have an inclination angle of 15 degrees, and the ratio of the paddle diameter to the diameter of the stirring cylinder is $\frac{D}{T}=0.45$, and the stirring paddles are made of nylon material by 3D printing. The motor is axially fixed in the rigid body bracket 330mm from the desktop through the fixture axial fixation, through the fixture of the lateral positioning nut to adjust the size of $r$ to indirectly control the position of the stirring shaft in the transparent cylinder (eccentricity $e$). The position of the acrylic transparent cylinder is fixed 90mm in front of the motor bracket, and the high-speed camera is fixed in front of the base to record the experimental process. The power was detected in real time using an AC/DC power meter (Yongpeng Company, Model: YP2012) with 0.2% precision. The sound decibel value is collected using the noise sensor developed by Shanghai SONBEST, and the data is output to the computer through the RS485 interface.

3.2 Test Methods

  1) Determination of mixing time

Water and glycerol were respectively added to the transparent cylinder to test the mixing characteristics of different viscous liquids. After the motor speed was increased to stable, 5g of red weak acid dye was quickly put in at the calibrated position as Fig. 3 shown. The more uniformly the dye was mixed, the darker the color was. The dye diffusion process would be recorded since the dye comes into full contact with the liquid. In order to avoid unavoidable factors such as reflections and shadows, the region "L" was selected as the monitoring area, where the slowest color mixing was observed. The gray scale value of the monitoring area was extracted based on the HSV color space conversion algorithm, and the process as Figure 5 shown. A black image was output when a fully red area is detected in the L region. Then, the L region image is binarized and the corresponding grayscale image and the grayscale area, which marked in red, close to about 80% of the average grayscale value are output. If the red area is greater than 80% and the subsequent average gray value stabilizes at 2% compared to before, it is considered that the time corresponding to the initial stable average gray value is the time required for uniform mixing.

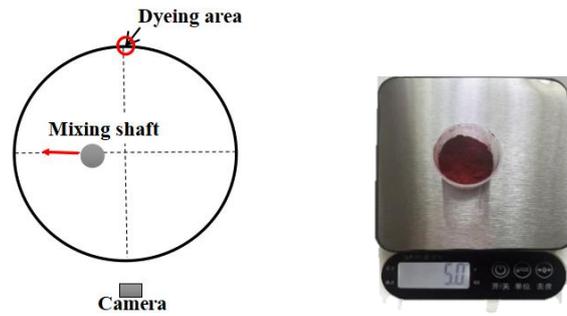

Fig. 3 Schematic diagram of the location of the dyestuff input port and the dyestuff map

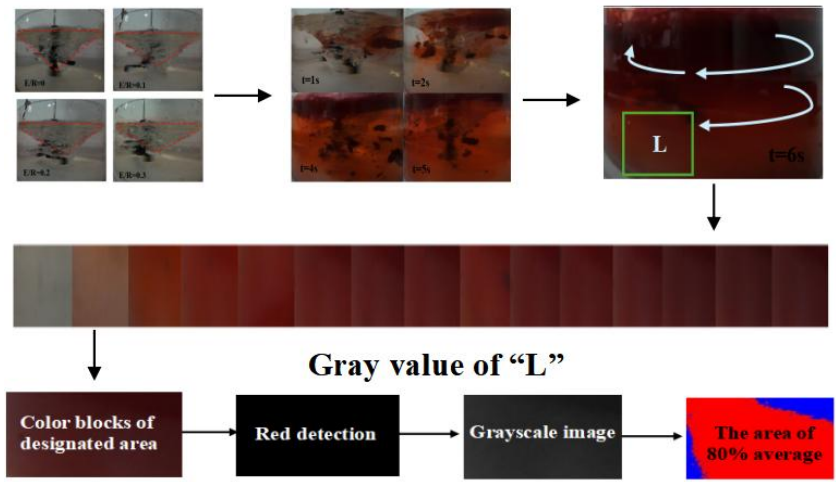

Fig. 4 Image processing process for the region "L"

2) Calculation of Lyapunov exponent

In this paper, the control variable method was used to study the law of the influence of eccentricity (0-0.5) and rotational speed (300-500rpm) on the level of chaos (LLE). A sound decibel collection device was constructed, see Fig. 1. Particle-particle and particle-wall collisions under solid-liquid mixing conditions were simulated by introducing heterogeneous particles with multiple scales (4mm/5mm/8mm) and different grammages (0.8g/0.2g/0.6g) in glycerol. The sound sources in the mixing process include particle-particle, wall paddle collision sound, glycerin slap sound, and multidimensional influencing factors.

The power and sound decibel data were sampled once per 1, and the baud rate of data acquisition was 9600 bps. The median filtering window Kernel_size was set to 5, and the median filter preprocessing of the time series was carried out to exclude interfering information. The effect of the noise reduction before and after the noise reduction is shown in Fig.5. Subsequently, the time delay $\tau$, embedding dimension $m$, and minimum time distance $d$ of the time series data were calculated. Finally, the calculation result of the parameter with the largest proportion in each item of data was taken for the subsequent LLE calculation.

1) Embedding dimension determination. The False Nearest Neighbors (FNN) method was applied, with a threshold set at 0.05 to construct $m_0$-dimensional the phase space of the dimensional time series. The distances between the time series points $x_i$ and at least 5 neighboring points $x_j$ in the $m_0+1$-dimensional embedding space were calculated according to the query algorithm, and the current minimum time distance and the proportion of false neighboring points were statistically calculated. When $m_0$ was increased to 10, the first false nearest neighbor ratio lower than 0.05 corresponds to the embedding dimension m, and the minimum time distance d of this group of data is the smallest among all time distances.

2) Time delay calculation. The mutual information method is adopted, and the np.histogram2d function is used to calculate the joint histogram between time series data slices. Lastly, the normalized probability distribution is used to calculate the mutual information MI. The time delay corresponding to the first time when MI reaches the minimum value is the final time delay.

The final calculation yields $m=5$, $\tau=1$ and $d=10$.

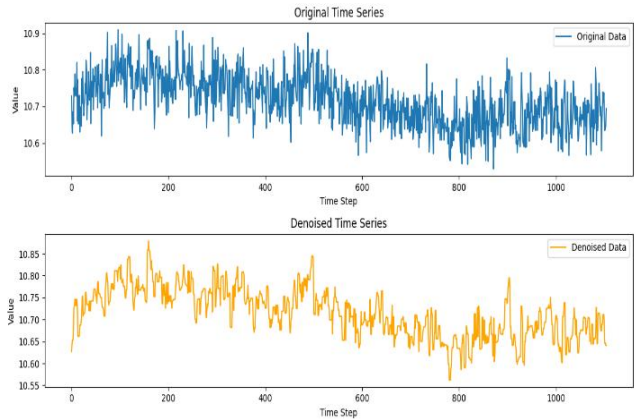

Fig. 5 Before and after noise reduction for a set of data

The value of k and the number of evolutionary steps s affect the expression of the final results on the dynamic characteristics of the system, as shown in Figure 7, the iterative step is taken as 50 to calculate the value of k for a set of values, and then according to the change of scattering to get a reasonable iterative step, and finally determine the k=5,s=30.

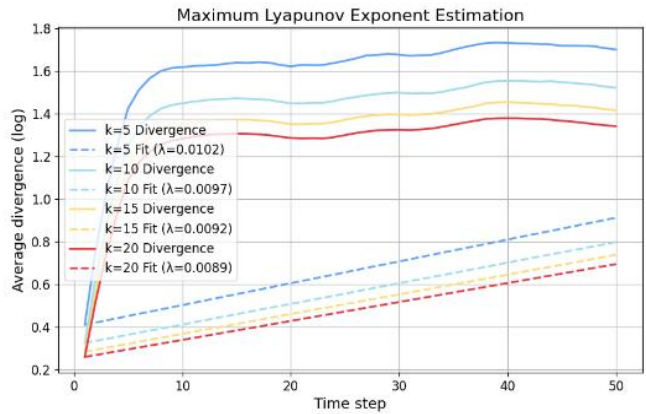

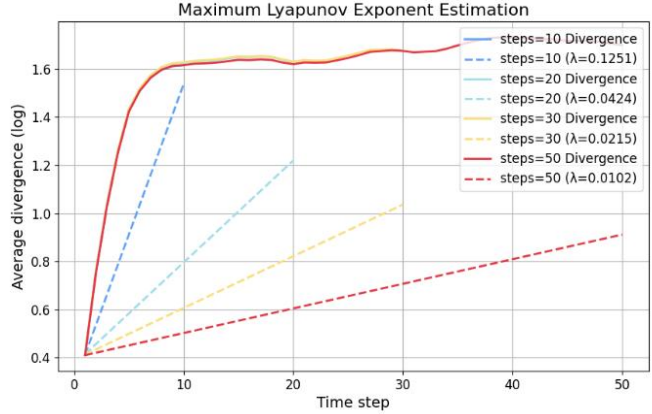

Fig. 6 Calculation of K-value and iteration step in Wolf's algorithm

## 4.Results and Discussion

4.1 Mixing characteristics

The performance in eccentric mixing of different dynamical viscosities liquids is shown in Fig. 8, and it is clearly observed that the vortex depth becomes smaller with the increase of eccentricity. The difference is that the vortex center position and vortex range do not change with the position of mixing shaft for liquids with almost no viscosity (water). However, for highly viscous liquids, due to viscosity and inertial force, the vortex center changes with the position of the mixing shaft, the vortex range decreases with the increase of eccentricity, the structure gradually becomes irregular, and even the vortex structure is broken up when the eccentricity reaches 0.5. Figure 9 shows the mixing process of dyes under different eccentricity. It can be observed that when dyes come into contact with any viscous liquid, they first cluster and start regular migration, then enter vortex spiral motion, and finally are thrown out of dispersion or continue the above vicious cycle.

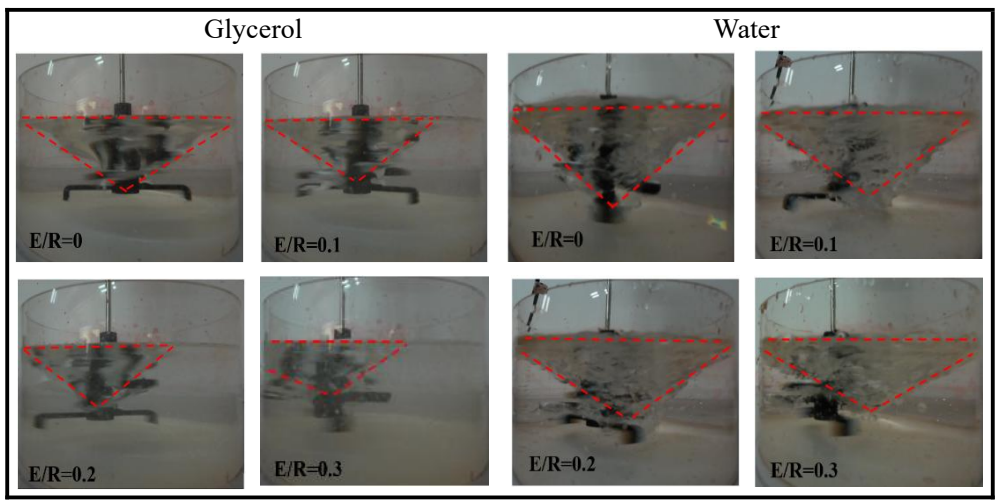

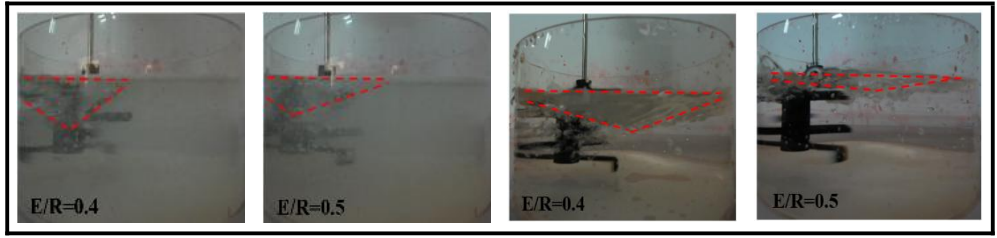

Fig. 7 Eccentric stirring vortex structure for liquids of different Viscosities

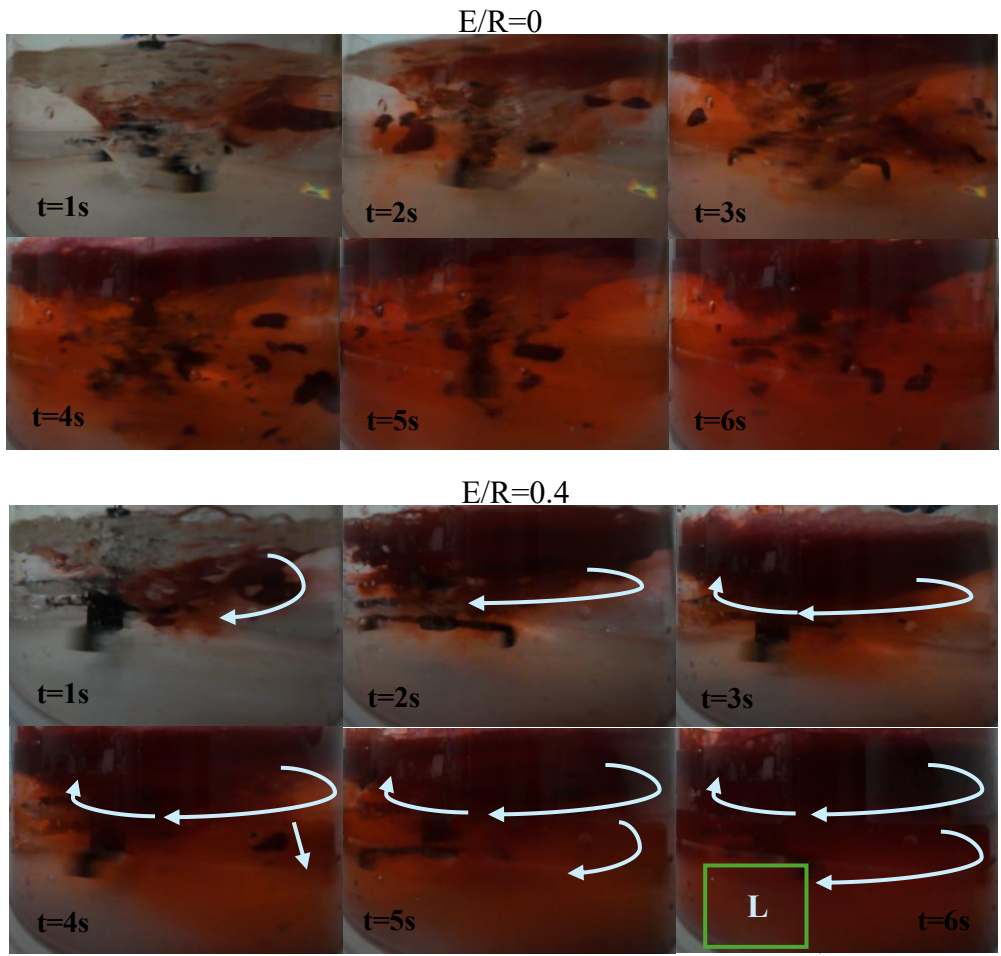

Fig. 8　Dye blending process in different eccentricity

Due to the clockwise rotation of the mixing shaft, the lower left region of the mixing cylinder, L (see Figure 9), is the slowest coloring, the most difficult to mix uniformly and is not subject to light reflection region. It is reasonable to select region L for mixing time detection. According to the image of region L, the result of the calculated gray value is shown in Fig. 11 and 10. Fig. 9 is the chromaticity bar of region L under a certain eccentricity, the more homogeneous the mixing, the darker the image color. According to Figs. 10 and 11, the mixing time tends to decrease and then increase with the eccentricity. When the rotational speed is constant and the eccentricity is 0.4, the time required for uniform mixing is the shortest, which is 40% shorter than that of the center stirring; when the eccentricity is controlled to be 0.4 and the rotational speed is 450 rpm, the mixing time is 48% shorter than that of the center stirring. It was analyzed that the larger the eccentricity of the stirring shaft, the shallower and smaller the vortex, the faster the solid dyes were released in the vortex and the degree of dispersion increased.

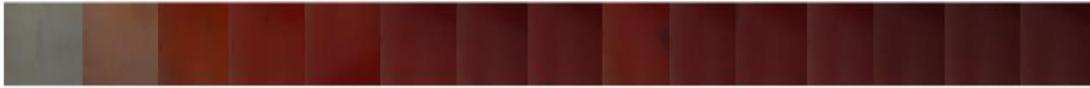

Fig. 9 Gradual increase of mixing time L-region chromaticity bar

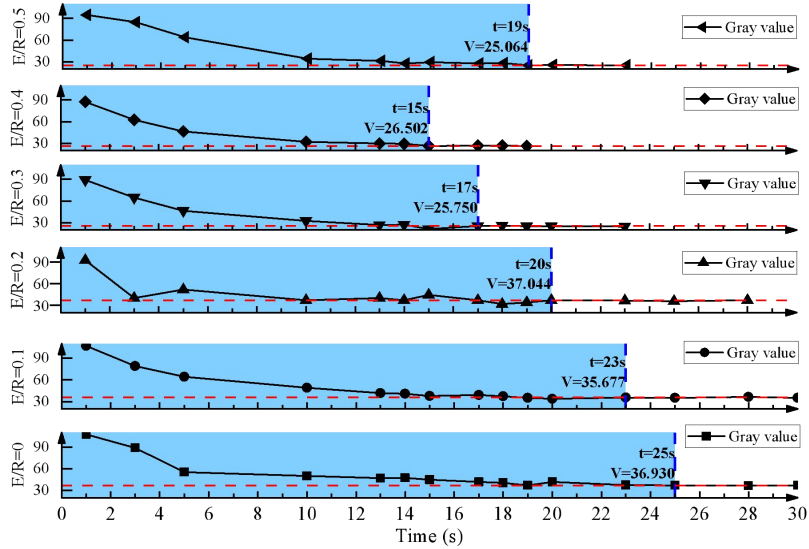

The variation of gray values with mixing time under different eccentricities

Fig. 10 Determination of mixing time for different eccentricity at 400 rpm.

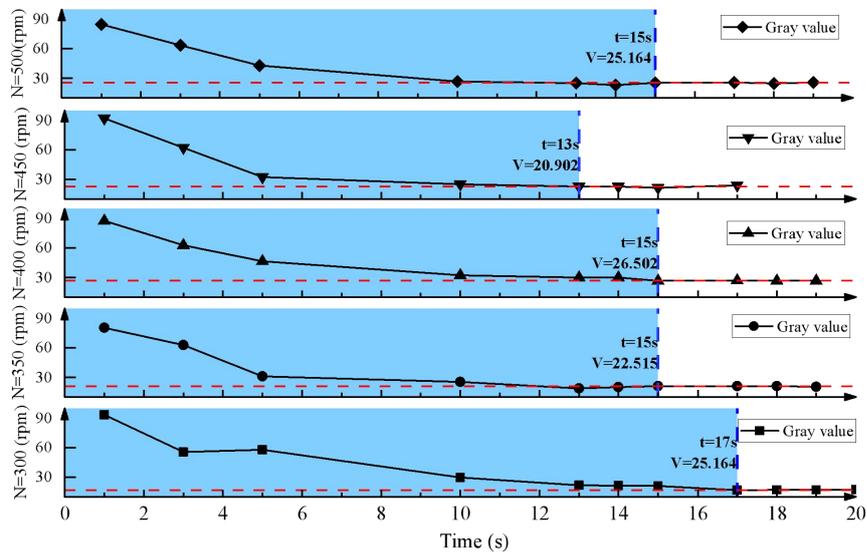

The variation of gray values with mixing time under different speeds

Fig. 11 Determination of mixing time for different rotational speeds with eccentricity of 0.4

4.2　Effect of eccentricity and rotational speed on LLE

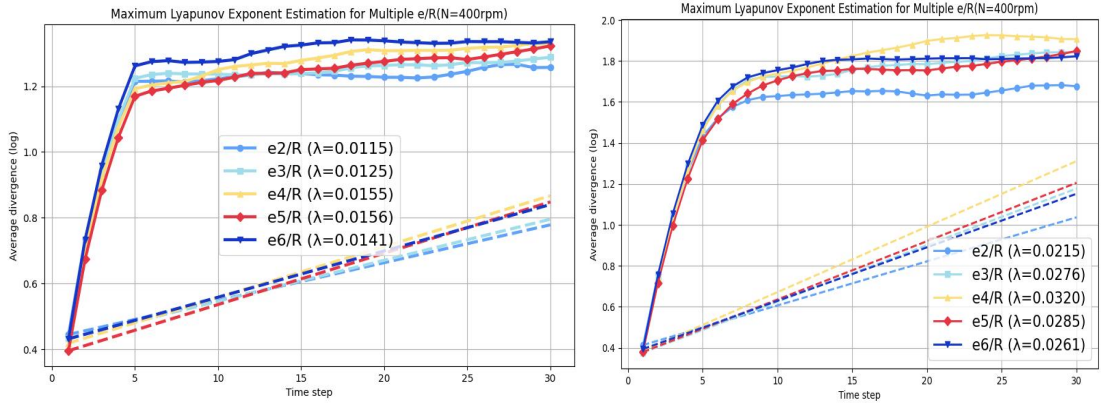

Fig. 12 LLE calculation results of power data with different eccentricity before and after noise reduction

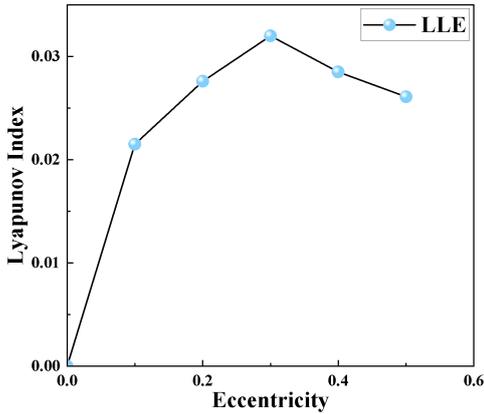

**The variation of LLE with different eccentricities**

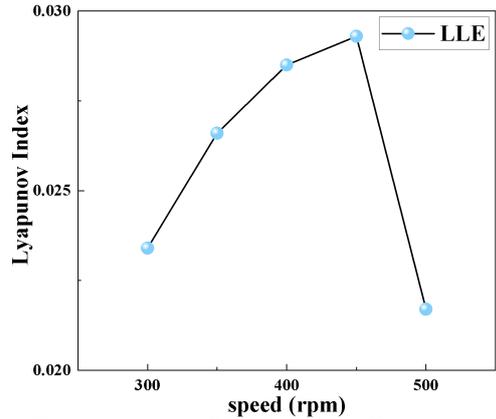

**The variation of LLE with different speeds**

Fig. 13 Variation of LLE with eccentricity speed after noise reduction

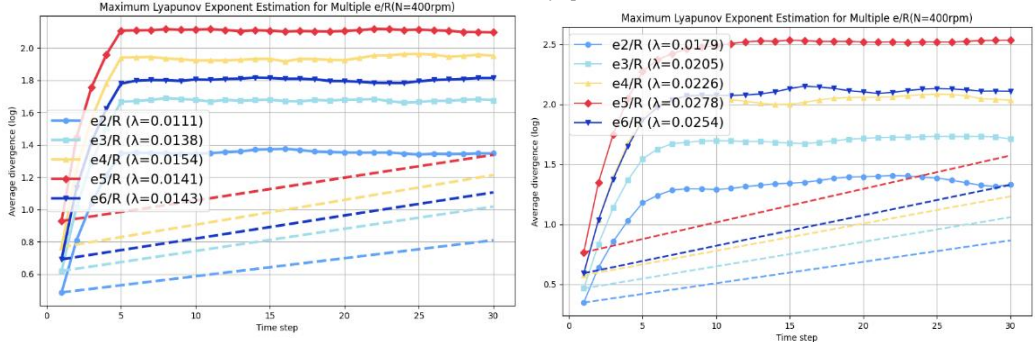

Fig. 14 Comparison of LLE before and after noise reduction of sound sequence data at N=400rpm

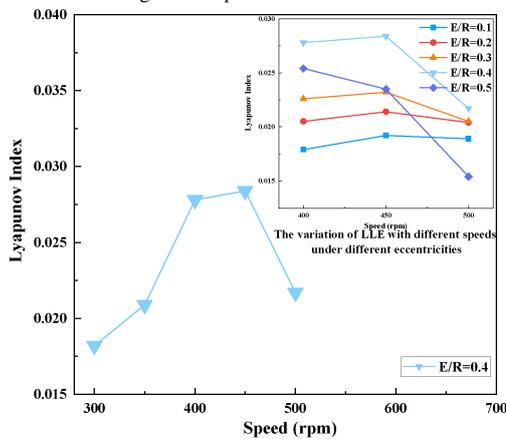

**The variation of LLE with different eccentricities**

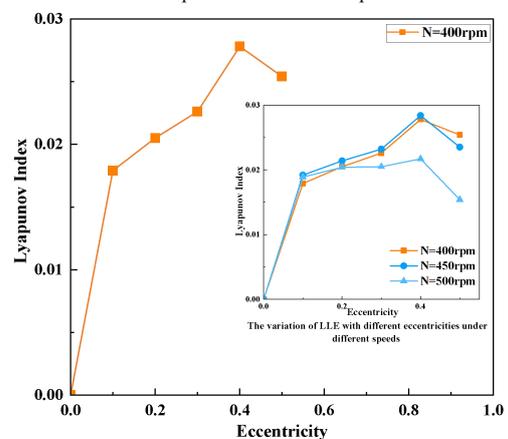

**The variation of LLE with different speeds**

Fig. 15 Variation of LLE with eccentricity, rotational speed

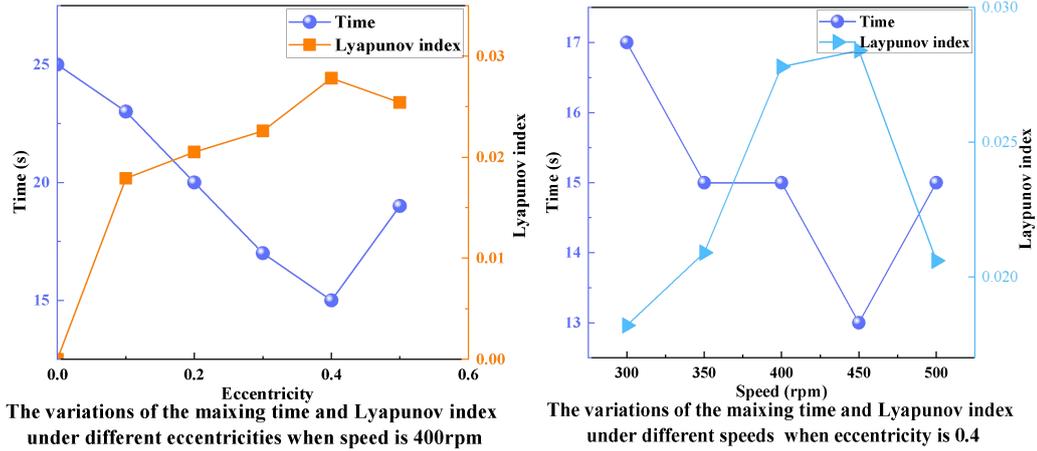

The variations of the maixing time and Lyapunov index under different eccentricities when speed is 400rpm

The variations of the maixing time and Lyapunov index under different speeds when eccentricity is 0.4

Fig. 16 Relationship between chaos index and mixing time for different eccentricity and rotational speeds

As shown in Figs. 12 and 13, the maximum value of LLE index before and after noise reduction of the power time series corresponds to different eccentricity, which means that both the value of the individual index itself and the overall trend are affected by the interference, which can be seen that there is a large amount of data interference in the power monitoring data, and the distribution of the data before noise reduction fails to reflect the systematic chaotic effect well. Before noise reduction, the size of the chaos index of the sound data shows waveform ups and downs with the change of eccentricity, and after noise reduction, it shows the trend of increasing and then rising, and the chaos index is the largest when the eccentricity is 0.4, which is in line with the trend of the mixing time, which indicates that the sound sequence filtering and noise reduction can better reflect the local chaotic characteristics of the mixing system.

According to Figs. 13, 15 and 16, it is found that the chaos index increases and then decreases with the increase of eccentricity and rotational speed under different control quantities, and the eccentricity, rotational speed and the degree of chaotic mixing are positively correlated if $\frac{E}{R} \leq 0.4$, $N \leq 400 \text{rpm}$. The chaos indicator peaks at $\frac{E}{R} = 0.4$, $N = 450 \text{rpm}$, which is 1.84 times the result of the minimum chaos indicator. The degree of chaotic mixing is negatively correlated with the mixing time, and the mixing time is shortest for the largest combination of chaos indicators, i.e., an eccentricity of 0.4 and a rotational speed of 450 rpm, which reduces the mixing time by 48% compared to the center mixing time.

4.3 Power Spectrum Comparison

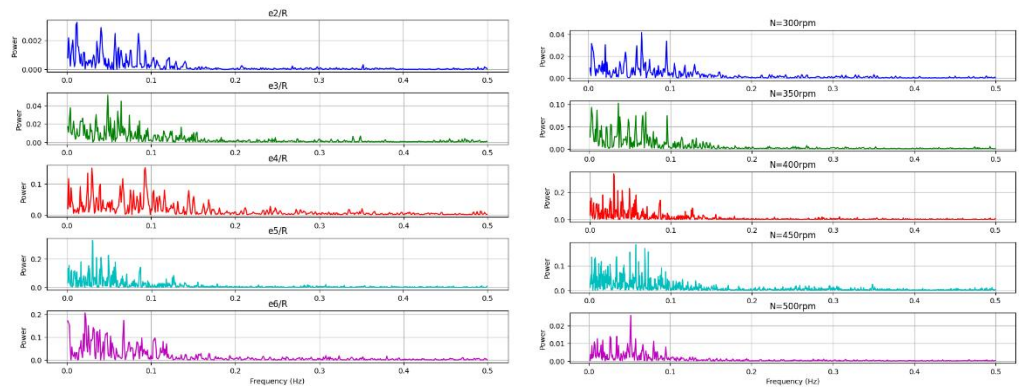

Fig. 17 Power spectrum of time series with different eccentricity (N=400rpm) and different rotational speeds (E/R=0.4)

The power spectrum demonstrates the oscillatory characteristics of the system at different frequency components of the time series. From Fig. 16, it is found that the power distribution at the low frequency range is intense and dense, the low frequency is related to the long-term trend or large-scale movement of the system, and the power density on both sides of the main peak of the power spectrum is asymmetric, and the peak density area is shifted in the low-frequency region, which is considered that the system's dynamic movement in different time scales has the characteristics of chaos, and it is considered that the sampling frequency is reasonable.

## 5. Relational equation study of LLE and eccentricity stirring parameters

According to the previous analysis, the chaos index (LLE) is directly related to the eccentricity and stirring speed, so $\lambda \propto e^{a1}, \lambda \propto N^{a2}$. In addition to this, the material viscosity, effective mixing diameter of mixing paddle, and other irregular spatial shapes can be the influencing factors for the variation of the chaotic mixing degree in the mixing tank. Considering the generalization and applicability of the formula without other irregular conditions are introduced, the eccentricity, rotational speed, material viscosity, and effective stirring diameter of the stirring paddle are utilized to dimensionless the chaotic mixing degree in the eccentric mixing.

If $M \propto \lambda$,

$$M = k(\frac{e}{R})^a (\frac{\rho ND^2}{\mu})^b$$

where k is the correlation coefficient, $\rho$ denotes the density of the liquid, D is the effective stirring diameter of the stirring paddle, $\mu$ is the viscosity of the liquid, e denotes the eccentricity of the stirring shaft, R is the radius of the stirring barrel, and N is the stirring speed.

K, a, and b in the model are global parameters that should be applied to all data points. Therefore, all data points were used to fit these three parameters. Taking the natural logarithm on both sides of the model transforms the equation into a multiple linear regression problem.

$$\ln M = \ln K + a\ln(\frac{e}{R}) + b\ln(\frac{\rho ND^2}{\mu}), (M > 0)$$

Let $y = \ln M$, $x_1 = \ln(\frac{e}{R})$, $x_2 = \ln(\frac{\rho ND^2}{\mu})$, $a = \beta_1$, $b = \beta_2$, and $\ln K$ is intercept, $\varepsilon$ is error term. And the experimental data were grouped using the least squares method to obtain 15 sets of data as shown in the table 2 below, and the data in the table were used to carry out multiple linear regression analysis to find the least squares solution to the equation.

Table 2 The result data of the regression equation

|          | x1=-2.302 | x1=-1.609 | x1=-1.204 | x1=-0.916 | x1=-0.693 |
|----------|-----------|-----------|-----------|-----------|-----------|
| x2=4.399 | y=-4.025  | y=-3.886  | y=-3.790  | y=-3.582  | y=-3.673  |
| x2=4.517 | y=-3.953  | y=-3.843  | y=-3.762  | y=-3.563  | y=-3.750  |
| x2=4.622 | y=-3.969  | y=-3.892  | y=-3.886  | y=-3.830  | y=-4.174  |

The matrix form of the equation can be obtained

$$Y = X \cdot \beta + \varepsilon$$

Where,

$$Y = \begin{bmatrix} y_1 \\ y_2 \\ \cdots \\ y_n \end{bmatrix}, X = \begin{bmatrix} 1 & x_{1,1} & x_{2,1} \\ 1 & x_{1,2} & x_{2,2} \\ \vdots & \vdots & \vdots \end{bmatrix}, \beta = \begin{bmatrix} \beta_0 \\ \beta_1 \\ \beta_2 \end{bmatrix}$$

So

$$\beta = (X^TX)^{-1}X^TY$$

The final least squares solution satisfying the equation is calculated as: a=-0.17,b=-3.14,K=22200.Then the dimensionless index M can be used to measure the degree of chaos in eccentric mixing.

$$M = k(\frac{e}{R})^{-0.17}(\frac{\rho ND^2}{\mu})^{-3.14}$$

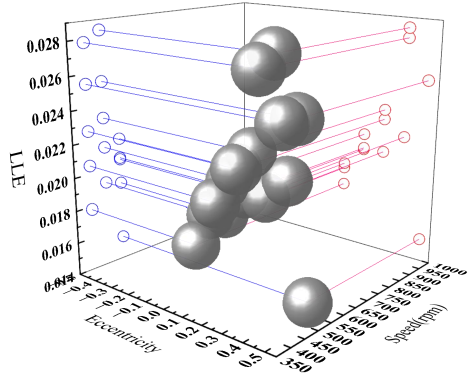

Fig. 18 Correlation analysis of the effect of eccentricity rate and rotational speed

The LLE size under arbitrary eccentricity and rotational speed is shown in Fig. 17, the LLE distribution in the region of rotational speed change is more balanced and dense, and the difference change is not obvious, while the LLE dispersion range is larger when the eccentricity is different, and the difference size range fluctuates, which means that in the case of controlling the variables, the change of eccentricity has a greater effect on the LLE than the change of rotational speed size, and the result is in line with the chaotic dimensionless results.

6.Conclusion

In this paper, the time nodes of solid-liquid mixing homogeneity during eccentric mixing were determined by combining tracer staining and image grayscale recognition techniques. It was found that when the eccentricity rate and rotational speed increased, the mixing time showed a trend of first decreasing and then increasing. The experiments showed that the optimal mixing efficiency was obtained when the eccentricity ratio E/R=0.4 and the rotational speed N=450 r/min. At this time, the mixing time was reduced by 48% compared to the center mixing time.

The analysis of the LLE calculation results reveals that the time series of sound decibel values can well reflect the local and long-term dynamic characteristics of the system.The eccentricity E/T and rotational speed N were changed respectively by using the control variable method. And it was proved that the LLE and the eccentricity and rotational speed showed a nonlinear relationship, and the highest degree of chaos was found when E/R=0.4 and N=450 rpm.

The dimensionless chaos index $M(M \propto \lambda)$ of the eccentric mixing system was obtained by using the mixture viscosity $\mu$, density $\rho$, effective mixing diameter $D$ of the mixing paddle, and eccentricity E/R. The dimensionless coefficients are obtained by solving for them using linear regression and least square method, and the result is $M = k(\frac{e}{R})^{-0.17}(\frac{\rho ND^2}{\mu})^{-3.14}$.